\documentclass[9pt,twocolumn,twoside]{opticajnl}
\journal{opticajournal} 

\setboolean{shortarticle}{true}
 \setboolean{memo}{true}

\doi{}
\dates{}

\title{\vspace{-1.5cm}Ultralow-power single-pass all-optical photon router}

\author[1,2]{J\'er\'emy Berroir}
\author[1,2]{Tridib Ray}
\author[1]{Alban Urvoy}
\author[1,*]{Julien Laurat}

\affil[1]{Laboratoire Kastler Brossel, Sorbonne Universit\'e, CNRS, ENS-Universit\'e PSL, Coll\`ege de France, 4 Place Jussieu, 75005 Paris, France}
\affil[2]{Welinq, 14 rue Jean Mac\'e, 75011 Paris, France}

\affil[*]{julien.laurat@sorbonne-universite.fr}

\begin{abstract} 
We demonstrate a novel approach for routing guided single photons by combining atom trapping and transverse light confinement with an optical nanofiber. The direction of photon propagation, either reflection or transmission, is controlled by a femtojoule-level beam.
\end{abstract}

\setboolean{displaycopyright}{false} 

\begin{document}
\maketitle

In the pursuit of optical networks and logic gates, there has been considerable interest in controlling the propagation of a light pulse using another optical beam \cite{Chai2017}. Most efforts have focused on optical switches, where the input light is either transmitted or absorbed depending on the control beam power. Achieving low-power operation involves enhancing optical nonlinearities, mostly through the use of cavity systems. For cavity-free, single-pass approach, tight transverse confinement of the beams is required. When combined with electromagnetically-induced transparency (EIT) in atomic gases \cite{Lukin}, such configurations have enabled efficient demonstrations, notably with hollow-core fibers \cite{Bajcsy2009}. In contrast, the realization of an all-optical router, capable of directing light into two distinct ports depending on the control, has been less explored, largely due to the increased complexity of achieving such functionality at low power.          

\begin{figure}[b!]
\centering
\includegraphics[width=0.96\linewidth]{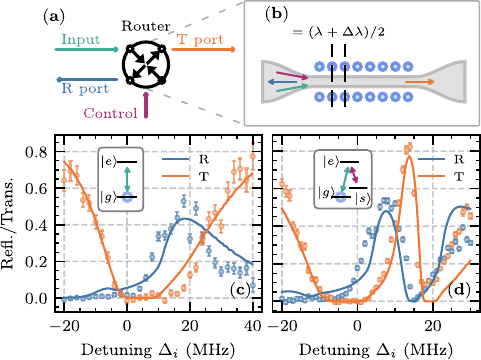}
\vspace{-0.15cm}
\caption{\small{(a) The input single photon is either reflected (R port) or transmitted (T port) depending on the control pulse state (on or off). (b) The router relies on arrays of cesium atoms trapped in the evanescent field of an optical nanofiber, with traps spaced at intervals nearly commensurate with the resonant wavelength. The nanofiber, drawn from a silica single-mode fiber, features a 1-cm-long tapered region with a 400-nm diameter, fabricated using the heat-and-pull technique. (c) Measured reflection and transmission for the input light without control, as a function of the input detuning $\Delta_i$. (d) Spectra with control light detuned by $\Delta_c=15$~MHz from resonance. The solid lines show the corresponding data from the theoretical model. Data points are averaged over 4000 experimental shots and the error bars correspond to the propagated Poissonian error of the photon counting probabilities.}}
\label{fig1}
\end{figure}

In this Memorandum, we report the realization of an all-guided, two-port optical router  for single photons operating at ultralow control power. Figures \ref{fig1}(a) and \ref{fig1}(b) present a schematic of the routing concept and its implementation. The device is built on a standard silica single-mode fiber, in which we created locally a 1-cm-long waist region with a 400-nm diameter \cite{Nieddu2016}. This nanofiber section provides strong transverse confinement of the guided light fields, while maintaining a transmission efficiency above 98\%. Arrays of individual cesium atoms are trapped in the evanescent field around the nanofiber using a two-color optical dipole trap that combines attractive red-detuned and repulsive blue-detuned lights \cite{Vetsch2010}. This platform has previously enabled a variety of demonstrations, including all-fibered EIT-based optical memories \cite{Gouraud2015,Sayrin2015} and single-photon generation \cite{Corzo2019}. In addition, dipole trapping the atoms offer the capability to obtain a lattice nearly commensurate with the resonant wavelength $\lambda$. This configuration enabled the observation of a large Bragg reflection \cite{Corzo2016}. Our work combines for the first time these two key features, EIT and atomic Bragg reflection, in an all-fibered configuration, opening the path to novel all-optical routing capabilities.

Specifically, the red-detuned trapping wavelength is offset by $\Delta\lambda=0.15$ nm from the exact Bragg configuration. This value is chosen to be close to resonance but still with the capability to trap atoms, and results in a slight shift of the Bragg condition out of resonance. Each red-detuned beam has a power of about 1 $\mu$W, while each blue-detuned beam, at 686 nm, operates at 4 mW. The atoms are loaded into the evanescent dipole trap, which is continuously on, via a 100 ms magneto-optical trap phase, followed by 8 ms of polarization gradient cooling. The atoms are then pumped into the $F=3$ ground state by applying a 300~$\mu$s pumping pulse. Residual magnetic fields are cancelled. The number of trapped atoms is estimated to about 1600.

The input light on the $F=3$ to $F'=4$ transition ($D_2$ line), detuned by $\Delta_i$, and the control beam on the $F=4$ to $F'=4$ transition, detuned by $\Delta_c$, are phase locked. The two beams are first combined using a fibered beam splitter, and their output is directed to a wavemixer setup. There, dichroic mirrors and a pair of volume Bragg gratings enable to combine these beams and the trapping lights into the fiber. Using a 90/10 beamsplitter, we can separate the reflected light, which is then sent through a 80-MHz-bandwidth cavity to suppress residual trapping and control light. At the transmission port of the nanofiber, a similar filtering stage is used. This configuration allows operation at the single-photon level for the input signal. Transmitted and reflected light are detected via single-photon counting modules. 

Figures \ref{fig1}(c) and \ref{fig1}(d) show the measured transmission and reflection as functions of the input detuning $\Delta_i$, with and without control beam. The input gaussian pulses have 300-ns FWHM with a mean photon number of 1, while the overlapping square control pulses have a duration of 1.4 $\mu$s. A reference pulse taken without atoms is used to calibrate the system transmission. For the reflection path, losses are carefully calibrated at each detuning as they are frequency dependent. In the absence of the control pulse, we observe the expected strong reflection, corresponding to the appearance of a bandgap, slightly shifted from resonance due to the small trap wavelength mismatch. The solid lines represent theoretical simulations based on the transfer matrix formalism for 3-level atoms \cite{Witthaut2010}, considering 1600 atoms and disorder induced by a filling fraction of the sites of 0.26. When the control pulse is present, we observe the opening of two bandgaps in the reflection spectrum \cite{Petrosyan2007,Schilke2012}, the new one being narrower and at the two-photon resonance. We optimized the control detuning such that the EIT transparency window opens at the maximum of the reflection and found the optimal value to be $\Delta_c=15$ MHz, as used in Fig. \ref{fig1}(d).

In this configuration, a specific input detuning allows for large reflection and near-zero transmission in the absence of control. When the control is applied, the reflection is suppressed and the transmission will be large. Due to the small mismatch from the Bragg condition, the maximum contrast for each channel occurs at a slightly different input detunings, namely $\Delta_i=14$~MHz for the reflection port and 15~MHz for the transmission port. We first characterize each channel independently. 

\begin{figure}[t!]
\centering
\includegraphics[width=0.96\linewidth]{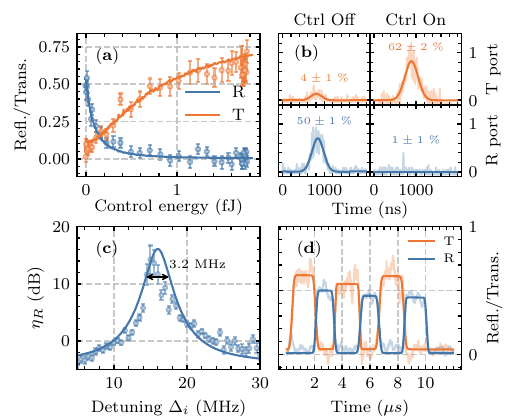}
\vspace{-0.15cm}
\caption{\small{(a) Measured reflection and transmission as a function of the control pulse energy. (b) Router truth table. Detection events in reflection and transmission are given without and with the control pulse. (c) Extinction ratio of the reflection path as a function of the input detuning. (d) Detection events in reflection and transmission at $\Delta_i=14.7$~MHz, as a function of time, with and without the control. Solid lines are gaussian fits in (b) and guides to the eye in (d). }}
\label{fig2}
\end{figure}

Figure \ref{fig2} summarizes the router's performance. In Fig. \ref{fig2}(a), we plot the reflection and transmission as a function of the control pulse energy. The data are fitted with exponential functions to extract characteristic switching energies. We find a value of 0.7~$\pm$~0.1 fJ for the transmission port and  0.16 $\pm$ 0.01 fJ for the reflection port. The difference stems from the cooperative nature of the Bragg reflection. The transmission is comparable to the best demonstrated platforms for single-pass all-optical switching \cite{Ruan2022}, with the added functionality of routing. Figure \ref{fig2}(b) presents the router's truth table, showing the transmitted and reflected pulses for the two control states, thereby demonstrating the binary behavior. An additional figure of merit is the extinction ratio and associated bandwidth for each output path \cite{Ruan2022}. Figure \ref{fig2}(c) shows this ratio for the reflection, along with a lorentzian fit to the data. We observe a maximum extinction ratio of 17 dB and a 3-dB-bandwidth of 3.2~MHz. For the transmission port, a ratio of 12 dB is measured. Finally, Fig. \ref{fig2}(d) illustrates the dynamic operation of the router using the same compromise input detuning of $\Delta_i=14.7$~MHz. As expected, a slight reduction in efficiency and contrast is observed compared to the previous individual cases. 

Beyond its fundamental significance as a first demonstration combining EIT and atomic Bragg reflection in a waveguide system, this work validates the use of a nanoscale waveguide to realize an all-guided, single-pass, two-port router operating at ultralow control power. Pushing performance to the single-photon-level control in this cavity-free regime is a long-standing challenge \cite{Chang}, with key applications for quantum technologies, and our implementation represents a significant step in this endeavor.\\

\noindent\textbf{Funding.} European Union Horizon 2020 (project DAALI No. 899275); Agence Nationale de la Recherche (projects 1DOrder ANR-22-CE47-0011 and QMemo ANR-22-PETQ-0010); Institut Universitaire de France.\\
\textbf{Disclosures.} The authors declare no conflicts of interest.\\
\textbf{Data availability.} Data available upon reasonable request.

\end{document}